\documentclass[12pt]{article}
\usepackage{amssymb}
\usepackage{amsfonts}
\usepackage{color}
\usepackage{amsmath}
\usepackage[margin = 1in]{geometry}
\usepackage{natbib}
\usepackage[doublespacing]{setspace}
\usepackage{amsthm}
\usepackage[flushleft]{threeparttable}
\usepackage{adjustbox}
\usepackage{caption}
\usepackage{subcaption}
\usepackage{hyperref}
\usepackage{graphicx}
\usepackage{multirow}
\usepackage{amsmath}
\usepackage{ulem}
\usepackage[table]{xcolor}
\usepackage{adjustbox}
\usepackage[toc,page]{appendix}
\usepackage{tabularx}
\usepackage{rotating}
\usepackage{pdflscape}
\usepackage{ulem}
\theoremstyle{definition}

\begin{document}

\title{Mass Shootings, Community Mobility, and the Relocation of Economic Activity}
\author{Miguel Cuellar\thanks{%
Department of Management Sciences, Abilene Christian University, miguel.cuellar@acu.edu} \qquad Hyunseok Jung\thanks{%
Department of Economics, University of Arkansas, hj020@uark.edu}}
\date{January 2026}
\maketitle

\begin{abstract} \doublespacing

\thispagestyle{empty}
\noindent Using foot traffic data for over 153,000 points of interest (POIs) near the sites of 42 mass shootings (2018-2022, U.S.), we evaluate the spatial-temporal impact of these tragic
events on community mobility and relocation of 
economic activity. Visits to nearby POIs decrease, while farther away POIs experience increased foot traffic, implying that communities shift their activities away from the shooting sites. The impact is stronger when stronger trauma responses are expected. Our results suggest that
mass shootings drive significant spatial-temporal displacements of economic activity and can lead to welfare losses due to distortions in optimal choices of time and location.
\medskip \bigskip


\end{abstract}

\newpage

\setcounter{page}{1} \renewcommand{\thepage}{\arabic{page}} %
\renewcommand{\thefootnote}{\arabic{footnote}}

\section{ Introduction}
Mass shootings have a profound effect on survivors, witnesses, and communities that experience them. The substantial loss of innocent lives is tragic on its own and clearly demonstrates that these events come at a great cost. However, the broader literature in economics, psychology, and beyond suggests that loss of life does not constitute the sole effect of mass shootings. Rather, it is accompanied by a variety of consequences on various fronts, including: worsening mental health for both direct victims and the surrounding community \citep{LoweGalea2017,SharkeyShen2021,WozniakEtAl2020}; exacerbation of political polarization \citep{DonohueEtAl2019,LucaEtAl2020,Yousaf2021}; and decreases in school enrollment and performance \citep{BelandKim2016,LevineMcKnight2024,BharadwajEtAl2021}.\footnote{For a recent survey of the literature on the causes and impacts of mass shootings, see \citet{Yousaf2023}.}

Although these tragic events are the subject of a growing number of papers, evaluations of their economic consequences are rare. 
\citet{BrodeurYousaf2022} is one of the few published studies that provide a comprehensive analysis of mass shootings from an economic perspective.  Using annual data, the authors find that mass shootings reduce the number of jobs and establishments in targeted counties by about 1.8\% and 1.3\%, respectively. They also find detrimental effects on earnings and housing prices in the years following the shootings.\footnote{Another important recent study, \citet{Cabral2025}, analyzes 32 non-mass shootings at public schools in Texas from 1995 to 2016, and finds exposure to a shooting harms human capital accumulation and leads to worse employment rates and earnings in the long term.}


In contrast to \citet{BrodeurYousaf2022}, which presents year-over-year and county-level effects, our analysis is centered on a more immediate and direct economic outcome: changes in visits to locations within five miles of a shooting site, right after the incident.  Specifically, this paper answers the following questions: Do mass shootings result in the displacement of economic activity, especially
toward locations farther from the shooting sites? Do mobility changes differ by location characteristics, such as substitutability? And are these changes related to the perceived risk of victimization or traumatic responses to the event? By answering these questions, our paper 
uncovers important mechanisms underlying the long-term, macroeconomic effects of mass shootings documented in \citet[][]{BrodeurYousaf2022}.

For our analysis, we use the mass shootings that occurred in the United States between 2018 and 2022, as documented by \texttt{Mother Jones}. 
This list excludes shootings from more conventionally motivated crimes, such as robbery or gang activities, and thus allows our analysis to be grounded on non-targeted and unpredictable events. To measure mobility, we use \texttt{Safegraph}'s ``weekly patterns" dataset, which includes foot traffic data and relevant characteristics for points of interest (POIs). Then, we employ event study and Difference-in-Differences (DiD) regressions to estimate the causal impact of mass shootings on community mobility over a period spanning ten weeks before and twenty weeks after the incidents.\footnote{ \citet{VM2025} employ a similar approach to estimate the effect of gun violence on community mobility. Their analysis, however, considers all gun violence incidents, including robberies and domestic violence, and uses foot traffic data aggregated for census tracts over a much shorter time period (28 days before and after a shooting).}

Our empirical findings can be summarized as follows. First, visits to nearby POIs decrease significantly  after a mass shooting, while farther away POIs experience increased foot traffic a few weeks later, suggesting spatial and temporal displacements of economic activity. 
For instance, in week 5 after a mass shooting, nearby POIs exhibit an average decline in visits of 26\% relative to the week before the shooting, while farther-away POIs show a 13\% increase 
in week 12.
Second, POIs that can be substituted with relative ease due to their business nature or the presence of a large number of competitors in the region 
exhibit more intensive changes. POIs that are 
difficult or undesirable to switch in the short run, such as healthcare facilities, 
show no 
sizable effect.\footnote{We discuss the welfare implications of the reallocation of economic activity in Section \ref{results-sub}. Even without measurable changes, welfare can still be affected. If the preference for the current location decreases due to a mass shooting but relocation is infeasible because of limited alternatives, a welfare loss may occur.} Finally, the mobility changes and relocation of economic activity are stronger when the shooting occurs in a previously safer area, when it receives more media coverage, and for POIs that are more heavily visited before the incident. Combined, the results suggest that people reallocate their activities across time and space due to these events,
especially under the conditions where stronger trauma responses are expected.

This paper contributes to the literature in several ways. First, it sheds light on an important driver for the long-term, macroeconomic effects of mass shootings 
found in other studies \citep[e.g.,][]{BrodeurYousaf2022}. After an event of this type, communities experience significant spatial-temporal displacements of economic activity,
likely incurring friction costs or efficiency losses due to distortions in optimal location and timing decisions. 
This can ultimately worsen various macro-level outcomes, including wages, employment rates, and housing prices. Second, our results reveal behavioral changes in response to trauma, linking the economic consequences to those documented in studies on trauma psychology  \citep{WozniakEtAl2020,SharkeyShen2021}, the economics of crime \citep{Braakmann2012,DustmannFasani2014}, criminology and sociology \citep{Garofalo1981,Hale1996}, and public health \citep{MarquetEtAl2020,GuiteEtAl2006}. Finally, this study finds that, in terms of per-POI impact, recreational spaces, places for temporary accommodation, educational facilities, especially primary and secondary, food and drink stores (e.g., grocery stores), restaurants, and bars are among the most impacted locations. This opens up an opportunity for future studies to use the variations in activities in these categories to identify other 
important relationships between economic variables.

The remainder of this paper is structured as follows. Section \ref{data} details the data sets used in this paper, with some descriptive statistics summarizing important mobility patterns in the data. Section \ref{strategy} describes our empirical strategy and regression models. Section \ref{results} presents the results, and Section \ref{conclusion} concludes. The online appendices include additional empirical results based on alternative specifications.

\section{ Data and Descriptive Analysis}\label{data}

\subsection{Mass shootings} \label{ms-data}
This study uses mass shooting data from  \texttt{Mother Jones},\footnote{\url{https://www.motherjones.com}}  a nonprofit newsroom producing investigative news on politics, criminal justice, and other fields. There are a few important differences between the \texttt{Mother Jones} mass shootings database and other lists of shootings used in the literature. First, elsewhere, following the FBI's definition, a mass shooting is usually defined as a shooting incident resulting in four or more deaths (not including the perpetrator), but since 2013 \texttt{Mother Jones}' reporting is based on a lower threshold of three or more deaths, following the mandate for federal investigation of mass shootings authorized by President Barack Obama. Second, \texttt{Mother Jones}' list excludes shootings from more conventionally motivated crimes, such as armed robbery, gang activity, and domestic violence.  Instead, the list includes incidents where the perpetrator's motivations are unclear or involve mental health, a history of violent behavior, among others.

This selection of mass shootings, therefore, allows our analysis to avoid mixing in the broader issue of gun violence and instead focus on non-targeted and unpredictable shootings, establishing reasonable grounds to treat these events as exogenous shocks. Exogenous shocks facilitate identification and, in the context of the present study, are \emph{ex ante} more likely to impact community mobility patterns by virtue of their unexpected nature.

Our analysis uses a total of 42 mass shootings that occurred between 2018 and 2022. 
The complete list of shootings, along with information on shooting date, number of deaths (excluding the perpetrator), and injuries, can be found in Online Appendix A. The human cost of the shootings is substantial, with 279 people dead and 302 experiencing injuries. The deadliest of these events took place at a Walmart in El Paso, TX and at the Robb Elementary School in Uvalde, TX - both resulting in over 20 fatalities.

\subsection{Community Mobility}

The foot traffic data used in this study comes from \texttt{Safegraph}'s weekly patterns dataset, which is based on a collection of cellphone location data from a large list of applications that send \texttt{pings} - that is, requests for GPS data - to the phone in which they are installed.\footnote{Papers using the Safegraph data include \citet{ChenRohla2018}, \citet{ChenEtAl2022}, \citet{ChenPope2020}.} \texttt{Safegraph} aggregates pings to produce a large panel-type dataset containing the number of weekly visits to distinct points of interest (POIs).\footnote{See \url{https://www.safegraph.com/guides/visit-attribution-white-paper} for more details on \texttt{Safegraph}'s methodology} In the dataset, a point of interest is any physical place of a non-residential nature, such as a restaurant, a museum, or a church, and it can belong to any of 180+ industry groups. In addition to tracking the number of visits to individual POIs within specified date ranges, the data include information that allows us to identify the locations and types of POIs: unique ID, latitude and longitude, location name, and North American Industry Classification System (NAICS) code. 

Our main analysis uses a total of 4.2 million observations for about 153 thousand distinct POIs located within five miles of the mass shooting sites, covering a period of 30 weeks for each event - 10 weeks before the shooting and 20 weeks after. Nonparametric estimation results, reported in Online Appendix B.2, support the choice of five-mile radius, where the impact of mass shootings on mobility vanishes 
beyond five miles from the shooting site.

Six shooting events took place in more than one location. These events, known as sprees, require a special handling since our analysis relies heavily on the distance between a POI and its associated shooting site. In these cases, an overall centroid may be used to represent the crime scenes, but we have opted to include all crime scenes as separate shooting sites, including all POIs within a five mile radius of each of the crime scenes, and assigning the minimum distance-to-shooting-site in the event of overlaps. This strategy improves precision and is also more adequate in our analysis given that, in some cases, the distance between crime scenes is substantial (e.g., the Atlanta Massage Parlor Shootings' victims were 30 miles apart). In Online Appendix A, events are individually identified as being sprees or having truncated time series.\footnote{
Our \texttt{Safegraph}  data start in January 2018 and end in December 2022, so the mobility data are truncated for some shooting events. For instance, the Marjorie Stoneman Douglas High School shooting, which occurred on February 14th, 2018, does not have a full set of 10 weeks of data before the event, and it is thus truncated on the left. Seven out of the 42 mass shootings have such truncated time series.}

\subsection{Supplemental Datasets}

Some additional datasets are used later in the paper to perform heterogeneity analysis. The first is a simple author-built media coverage dataset, capturing the number of event-related articles available online within one-week of each shooting. Articles are identified using broad internet search terms, such as ``Boulder, Colorado Shooting" or ``Tulsa, Oklahoma Shooting." The second dataset consists of county-level, time-adjusted and population-weighted statistics on homicide rates based on data from the Centers for Disease Control and Prevention.\footnote{\url{https://www.countyhealthrankings.org} (\texttt{County Health Rankings and Roadmaps}).}

\subsection{Descriptive Analysis}\label{DA}

\begin{figure}[h!]
	\caption{Descriptive Figures of Mobility Changes After Mass Shootings}
	\centering
	\begin{subfigure}{\linewidth}
            \caption{Changes in Visits to McDonalds Locations After Mercy Hospital Shooting}\label{mcdonalds}
            \centering
		\includegraphics[width=\textwidth]{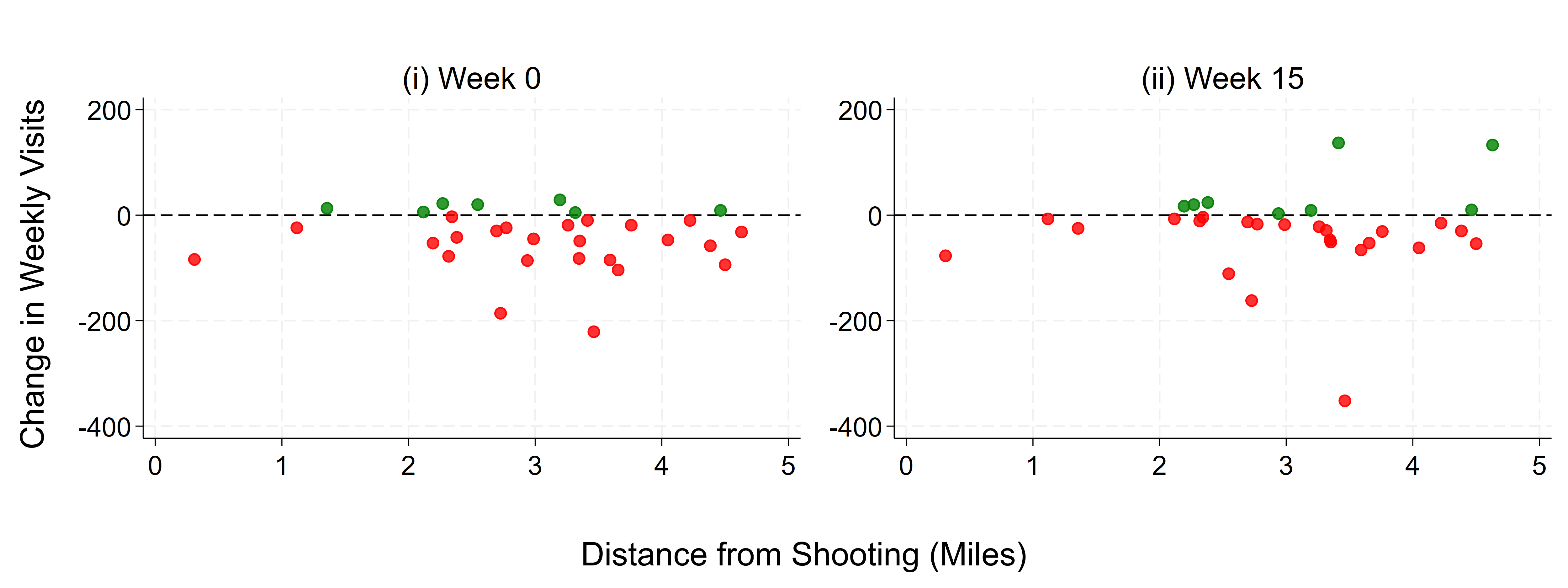}
	\end{subfigure}
        \par\bigskip
        \begin{subfigure}{\linewidth}
             \caption{Evolution of Aggregate Visits and Active POIs - By Distance Band}\label{pois_and_visits_quantile}
             \centering
            \includegraphics[width=\textwidth]{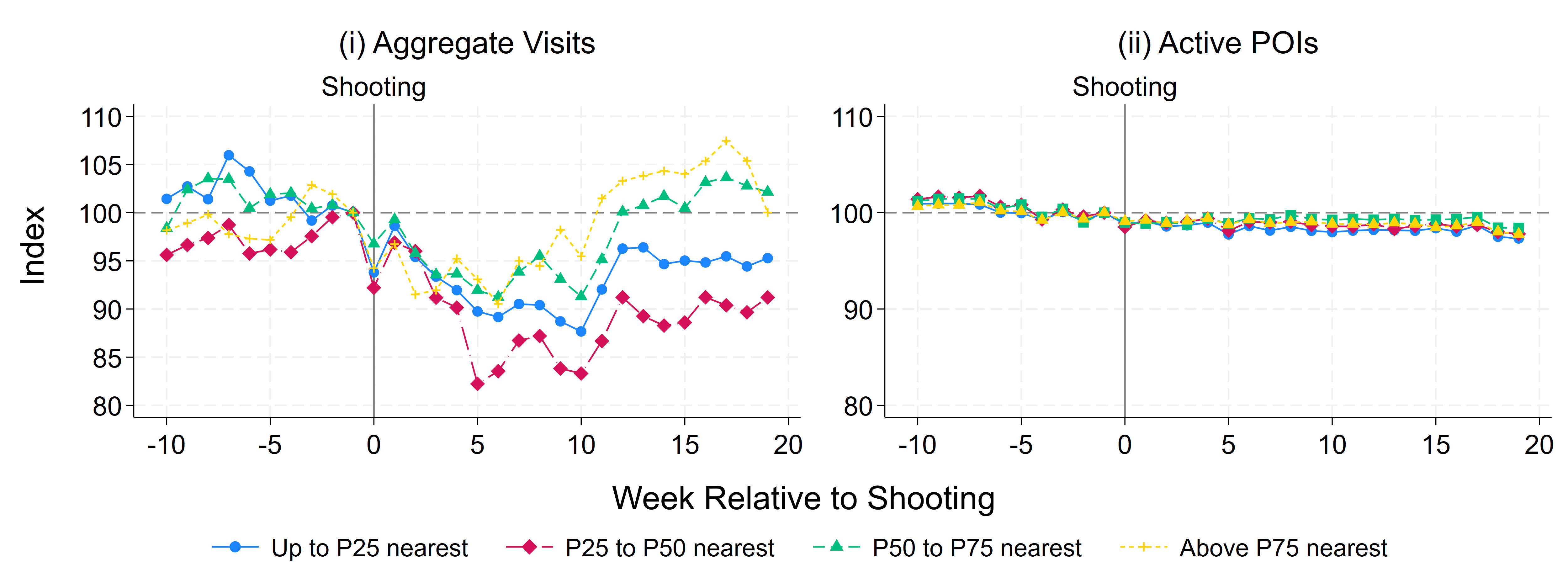}        
        \end{subfigure}
        \caption*{\footnotesize \textit{*Note:} Subfigure (b) plots the evolution of visits and active POIs, using the week immediately before the shooting as the baseline. Sprees and events with incomplete time series are excluded.}
\end{figure}

This section offers an introductory description of the mobility patterns observed in the data. Figure \ref{mcdonalds} shows the changes in weekly visits to \texttt{McDonald's} locations in Chicago after the Mercy Hospital shooting on November 19$^{th}$, 2018.\footnote{This exercise uses a franchise to minimize heterogeneity across locations. The y-axis represents the changes in weekly visits at each location compared to the week right before the shooting.}
 Clearly, most \texttt{McDonald's} locations experienced a decrease in weekly visits in both week 1 and week 15 after the shooting (red dots), but we do see some locations farther from the shooting site experienced an increase in visits in the latter period (green dots). This may suggest that farther away locations recover more quickly after the shooting, or even that a substitution mechanism is at play. That is, customers now prefer farther away locations rather than those in close proximity to the site of the incident.

We then compute an index measure of weekly \textit{aggregate} visits for all non-truncated, non-spree shooting events as a preliminary exercise. The index takes a value of 100 for the week prior to the shooting (week -1) and is calculated 
separately by distance band. Distance bands are concentric zones surrounding a shooting site, with cutoffs determined by the quantiles of the distances between POIs and the shooting site. For instance, the first quartile distance band comprises POIs located within the bottom 25\% of the distances, while the second quartile distance band includes POIs located between the 25th and 50th percentiles of the distances. Distance quantiles are computed at the shooting event level, and thus the physical sizes of the distance bands can differ from place to place.\footnote{The median POI distance for each shooting event is reported in Online Appendix A. The full relationship between quantile distance and physical distance (in miles) for each shooting event is shown in Figure C.1 in Online Appendix C.}

We prefer quantiles over raw physical distance (in miles) to define our distance bands since it accounts for differences in POI density in the various shooting areas and thus captures relative, locally ``perceived distance." This measure of perceived proximity can reflect relative accessibility in a local community, and therefore facilitates linking the observed mobility changes in our results to the behavioral changes or potential welfare losses among community members.\footnote{A potential concern is that if POIs are highly concentrated in a specific area, the distance bands based on quantiles can be physically too close, failing to represent distinct areas in a community. Figures C.1 and C.2 in Online Appendix C demonstrate that this is not a concern in our data. In these figures, we plot the 
relationship between raw distance and distance quantile, as well as the density of a given POI 
category at each distance quantile. Overall, 
POIs in the dataset are well spread across the areas surrounding the shooting sites, and POI 
categories are also relatively uniformly distributed.}



Figure \ref{pois_and_visits_quantile} (i) displays a substantial reduction in aggregate visits beginning the week of the shooting, across all distance bands. A faster recovery is observed for the POIs in the third and fourth quartile distance bands, and recovery is not complete for the first and second quartile distance bands even after 20 weeks. This aligns with the patterns observed in Figure \ref{mcdonalds}. Visits to the POIs in the third and fourth quartile distance bands rise even above the pre-shooting levels, suggesting potential spatial reallocation. 
Another interesting pattern is that the POIs in the second distance band exhibit a larger decline in weekly visits than those in the first distance band. To some extent, this may be due to the presence of authorities, visitors, or the media near the shooting sites.

Figure 
\ref{pois_and_visits_quantile} (ii) plots a similar index for active POIs, which displays the evolution of the number of POIs for which nonzero visits are reported in the data. The time-evolution of this index is much more subtle, indicating that the changes in mobility patterns are neither driven by POI closures, nor merely by missing some of the POIs during the period of analysis.

\section{Empirical Strategy}\label{strategy}

Our empirical strategy for identifying the causal impact of mass shootings on community mobility and relocation of economic activity hinges on the assumption that there is no fundamental change in community mobility 
prior to the week of mass shooting. We argue that this is a reasonable assumption given that our analysis focuses on non-targeted and unpredictable shootings, as discussed in Section \ref{ms-data}. Also, the pre-trends observed in Figure 
\ref{pois_and_visits_quantile} do not show any notable changes near the week of mass shooting, supporting the assumption.

We use the following event study design to estimate the causal impact of mass shootings:

\begin{eqnarray} \label{sp}
\quad y_{i,t} &=& \sum_{b \in \mathcal{B}}^{} \sum_{\ell = -10}^{\ell = -2} \beta_{b,\ell} D_{i,t}^{b,\ell} + \sum_{b \in \mathcal{B}}^{} \sum_{\ell = 0}^{\ell = 19} \beta_{b,\ell} D_{i,t}^{b,\ell} + \delta_i + \gamma_t + \epsilon_{i,t},
\end{eqnarray}
where $y_{i,t}$ is weekly visits to POI $i$ in year-week $t$, the set $\mathcal{B}$ includes distance bands,\footnote{See Section \ref{DA} for the definition of our distance bands. In our analysis, we use either two-quantile or four-quantile distance bands. Online Appendix B.2 reports results estimating the impact on mobility as a function of continuous distance. The overall patterns are consistent with those based on distance bands.} and $\ell$ is the week number relative to shooting omitting the week right before the shooting (baseline week). The dummy $D_{i,t}^{b,\ell} = \textbf{1}\{ B_i = b, t - ST_i = \ell \}$, where $B_i$ is the distance band to which POI $i$ belongs, and $ST_i$ is the year-week of the mass shooting associated with POI $i$. Therefore, $D_{i,t}^{b,\ell}$ is an indicator for POI $i$ being in distance band $b$ and relative week $\ell$, and thus $\beta_{b,\ell}$ calculates the average change in visits in distance band $b$ in week $\ell$ 
relative to the week right before the shooting.
The specification includes POI fixed effects, $\delta_i$, and year-by-week fixed effects, $\gamma_t$, controlling for potential business cycles at local-levels.\footnote{Due to the large number of POIs in the data (approximately 153,000 distinct locations), accounting for POI fixed effects while estimating Equation (\ref{sp}) is computationally challenging. In our case, it created some computation issues. Therefore, to facilitate estimation, we demean weekly visits by POI and estimate Model (\ref{sp}) without including POI fixed effects. We also considered alternative specifications with coarser time fixed effects, such as year–by-month effects, and the results were qualitatively unchanged.} The standard errors are clustered at the POI level.

We also compute DiD-style estimates that compare the magnitude of the impact across different distance bands
by reparameterizing Model (\ref{sp}) such that

\begin{eqnarray}\label{sp2}
\quad y_{i,t} &=& \sum_{\ell \in \mathcal{G}}^{} \beta_{\ell} D_{i,t}^{\ell} + \sum_{b \in \mathcal{B}/\{b^{\ast}\}}^{} \sum_{\ell \in \mathcal{G}}^{} \beta_{b,\ell} D_{i,t}^{b,\ell} + \delta_{i} + \gamma_{t} + \epsilon_{i,t},
\end{eqnarray}
where $\mathcal{G} = \{-10,-9,\ldots,19\}/ \{-1\}$, $D_{i,t}^{\ell} = \textbf{1}\{t - ST_i = \ell \}$, $b^{\ast}$ is the reference band and thus $\beta_{b,\ell}$ in Model (\ref{sp2}) calculates the impact relative to that on the reference.
\color{black}

\section{Results}\label{results}

\subsection{Mobility Patterns after Mass Shooting}\label{results-sub}

\begin{figure}[h!]
\caption{Impact of Mass Shootings on Community Mobility}
    \begin{subfigure}{\linewidth}
        \caption{Event-Study Estimates}\label{noDiD_combined}
        \centering
            \includegraphics[width=0.8\textwidth]{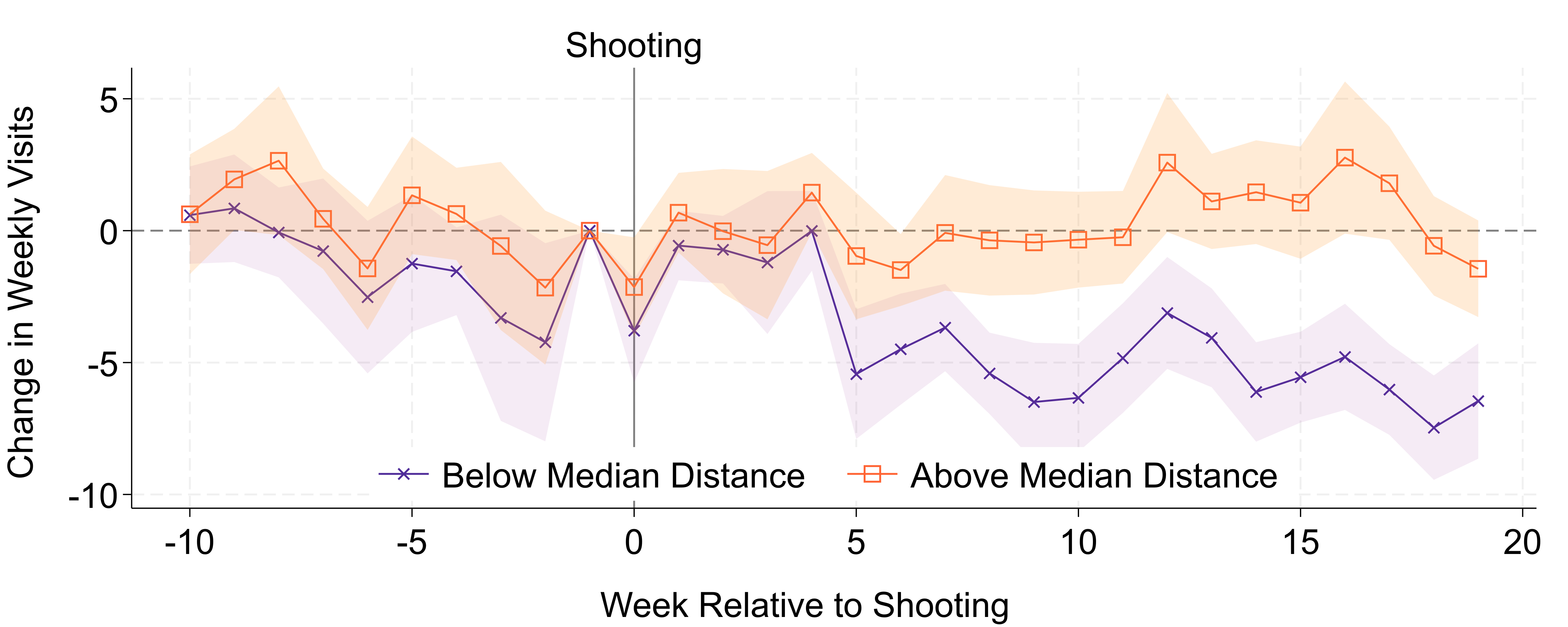}       
    \end{subfigure}
    \par\bigskip
    \begin{subfigure}{\linewidth}
        \caption{Difference-in-Differences Estimates}\label{DiD}
        \centering
            \includegraphics[width=0.8\textwidth]{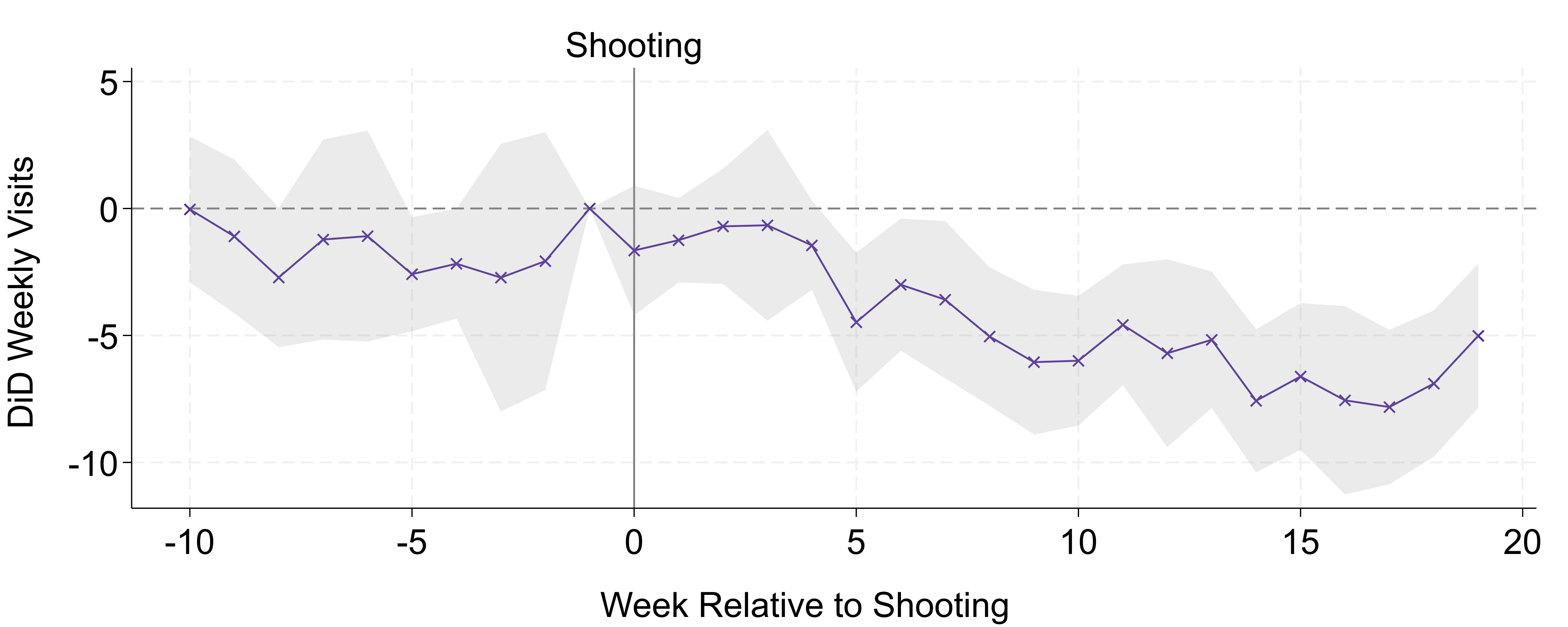} 
    \end{subfigure}
    \caption*{\footnotesize \textit{*Note:} 95\% confidence intervals are shown. Standard errors are clustered at the POI level. In subfigure (b), the reference group is the POIs above median distance from the shooting site.}
\end{figure}

Figure \ref{noDiD_combined} plots event study estimates obtained from Model (\ref{sp}) for two distance bands split across median distance from the shooting site.  POIs in both distance bands experience a significant decline in weekly visits immediately following the shooting (week 0). Visits then return to pre-event levels and remain at those levels through week 4. This pattern differs from that observed in Figure \ref{pois_and_visits_quantile} (i). One possible explanation for this discrepancy is 
that the mobility patterns in Figure \ref{pois_and_visits_quantile} (i) are computed by aggregating visits across all POIs in the sample, whereas the event study estimates in this section capture per-POI effects. Therefore, the discrepancy may indicate that in the early period, the impact was concentrated on a small number of POIs, possibly those with high visit density, and then spread more broadly across space in later periods, leading to a large decline in both aggregate and per-POI levels.


After week 4, POIs near shooting sites experience sizable, statistically significant declines in weekly visits for the rest of the period. This pattern is robust to an alternative specification using four distance bands instead of two (Online Appendix B.1), where, similar to the patterns in Figure \ref{pois_and_visits_quantile} (i), the declines in visits within the 1$^{st}$ quartile distance band are small relative to those in the 2$^{nd}$ quartile distance band, which may reflect the presence of authorities, visitors, or the media near the shooting sites.

POIs above the median distance, on the other hand, maintain visits at pre-event levels through week 11 and even exhibit a meaningful increase in visits for much of the remaining period. Two potential factors may explain the increase in weekly visits to these POIs. First, the resumption of previously postponed economic activity by individuals who normally use those POIs. Second, the relocation of activities to further away places by individuals who, prior to the event, typically used the POIs near the shooting site. The former implies only a temporal displacement of activities, but the latter suggests both temporal and spatial displacements. 


Given the small initial decline in visits to faraway POIs and then a notable, persistent increase later in the period, the resumption of postponed activities is unlikely to fully explain this rise, suggesting some degree of substitution or relocation of economic activity across regions.
That is, visits lost from the POIs in close proximity to the crime scene might have been redistributed, to some extent, to the POIs far from it. We further explore this relocation channel in Section \ref{hetero} by analyzing the impact by POI type and POI substitutability.


This substitution effect, however, does not kick in immediately after the shooting. Instead, 
several weeks go by before a significant divergence in post-shooting weekly visit trends is observed across regions. This may be due to the difficulty of substitution in the presence of search and switching costs \citep{Klemperer1987,Wilson2012,Wezsacker1984}. Therefore, the relocation of economic activity implies friction costs or efficiency losses due to a deviation from the optimal choices, which, in turn, could affect more macro-level economic outcomes such as wage, unemployment rate, and housing price, among others \citep[\emph{a la}][]{BrodeurYousaf2022}.\footnote{Compared to mere temporal displacement, the relocation of activities can result in more enduring and persistent social costs since the new location is likely sub-optimal compared to the previous one.}

Note that the changes in weekly visits in Figure \ref{noDiD_combined} are ``per POI"; therefore, the aggregate impact can be substantial. Since there are over 150 thousand POIs in the data, the overall impact will be the product of the estimated change per POI and the number of POIs in a given distance band. For instance, the reduction of about 5.4 visits per POI in week 5 for those POIs below the median distance will amount to an overall decrease of over four hundred thousand weekly visits. 
The median value of nonzero weekly visits to nearby POIs in the dataset during the week prior to the shootings (the baseline week) is 21, and thus a decrease of 5.4 visits constitutes roughly a 26\% decline relative to the median.

The differential mobility patterns in the two regions persist over time, resulting in a significant gap in weekly visits. Figure \ref{DiD} plots the DiD estimates for POIs below the median distance (i.e., nearby POIs) with the reference group being the POIs above the median distance (i.e., far away POIs). The graph thus directly quantifies the gap in weekly visits over time between the two regions and its significance. The gap becomes statistically significant after week 4 and increases to about 8 visits in week 17. The same pattern is observed from an alternative specification with four distance bands, although estimates for first-distance-quartile POIs lack precision (see Online Appendix B.1).

Our results in this section are robust to the thresholds used to define distance bands. Online Appendix B.2 provides nonparametric results based on splines (piece-wise polynomials), where we estimate the evolution of weekly visits as a function of ``continuous" distance. The patterns are consistent with those reported in this section. The spline results also support setting 5 miles as the cutoff for sample selection, as estimates approach zero. 

\subsection{Evidence for Relocation of Economic Activity} \label{hetero}

In this section, we analyze the impact of mass shooting by POI type and POI substitutability. If individuals relocate their activities from  POIs near the shooting site to those farther away as response to the tragic event, we would expect to see similar magnitudes of increases and decreases in weekly visits for the same type POIs across the two regions. Also, if relocation is the primary driver of the mobility changes observed in the previous section, we would expect a larger impact on the POIs that are easily substitutable.

\subsubsection{Impact by POI Type}

Changes in weekly visits are estimated by distance band and POI category, using two-quantile distance bands and 15 POI categories. 
POIs are classified based on NAICS codes.

\begin{figure}[h!]
    \centering
    \caption{Change In Weekly Visits by POI Category}
        \centering
        \label{zone_effect_spree}
        \includegraphics[width=\linewidth]{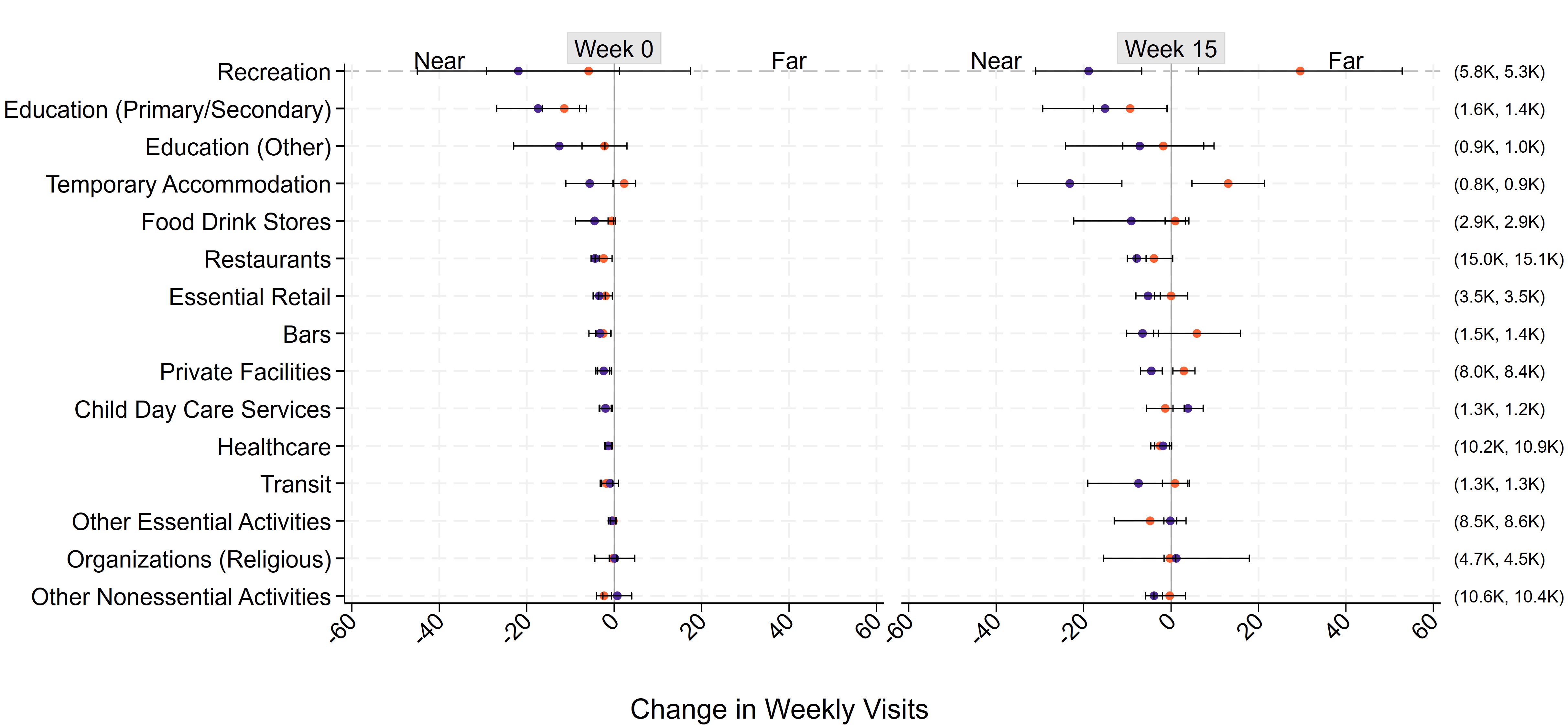}
    \caption*{\footnotesize \textit{*Note:} 95\% confidence intervals are shown. Standard errors are clustered at the POI level. ``Near''/``Far'' refers to POIs below/above median distance from the shooting location. The two numbers shown in parentheses on the right side of the figure correspond to the counts of nearby and far-away POIs, respectively, for each category.}
\end{figure}

Figure \ref{zone_effect_spree} reports the changes in visits and corresponding 95\% confidence intervals for various categories in the week immediately after the event (week 0) and approximately four months after the shooting (week 15).\footnote{Online Appendix D.1 includes the results for weeks 5 and 10.} The estimates indicate that decreases in visits in week 0 are quite general across a variety of POI types in both distance bands, especially closer to the shooting site, implying widespread economic costs of mass shootings.

Notably, for recreation, temporary accommodation, bars and private facilities (e.g., real estate, professional services), the increase in visits to farther-away POIs in week 15 closely mirrors the decrease to nearby POIs, pointing to potential relocation within these categories. Such relocation patterns are observed only for some selected categories, suggesting that POI characteristics may play a role in the substitution effects.

Heterogeneity of the impact across POI types is also notable. Recreation, education, temporary accommodation, food/drink stores exhibit larger changes, although their confidence intervals are also wider. The large and significant effect on education, which follows immediately after shootings and persists over time, deserves a look. For example, the decrease in visits to primary and secondary schools even in week 15 is estimated to be around 15.1 for nearby POIs and 9.3 for faraway POIs, which in aggregate amounts to a reduction of about 24 and 13 thousand visits, respectively. Disruption of educational activities not only hampers student learning, but also has negative effects on 
parents' economic activity, such as reduced work hours and productivity \citep{GarciaCowan2024}. Therefore, this large impact 
on education suggests a substantial and broad social cost.

\subsubsection{Impact by POI Substitutability}
The relative size of the changes in weekly visits across different POI types in Figure \ref{zone_effect_spree} suggests that ease of substitution might play a role in the observed patterns. It is relatively easy to change the recreational spaces you frequent and thus we see a significant shift in visits from nearby places to far-away places for recreation. In contrast, switching healthcare providers is difficult or undesirable in the short run, which accounts for the negligible effect on healthcare facilities four months after the shooting. 

\begin{figure}[h!]\caption{Change in Weekly Visits by Substitutability}
\centering
\par\bigskip
    \begin{subfigure}{\linewidth}
        \small
        \centering
        \renewcommand{\arraystretch}{1.75} 
        \caption{Measures of Substitutability}\label{substitutability_table}
        \let\center\empty
        \let\endcenter\relax
        \centering
        \resizebox{0.9\textwidth}{!}{\begin{tabular}{p{0.15\linewidth}p{0.4\linewidth}p{0.4\linewidth}p{0.1\linewidth}}
\hline
Measure &
  Description &
  High &
  Low \\ \hline
Common \quad Categories &
  Selecting a set of POI types based on what appears easier to substitute &
  \underline{POIs} of the following types: \emph{Food/Drink Stores, Gas Stations, Recreation, Restaurants/Bars, Retail, Temporary Accommodation} &
  Others \\
Competition Level &
  Selecting POIs based on number of competitors within 5 miles from the shooting site. Competitors are identified using 4-digit NAICS codes &
  \underline{POIs} that have above-median number of competitors in the region &
 Others \\
\hline
\end{tabular}}        
    \end{subfigure}
\par\bigskip
\par\bigskip
    \begin{subfigure}{0.9\linewidth}
    \centering
        \caption{Using Common Categories}\label{substitutability_cats}
            \centering
            \includegraphics[width=0.9\textwidth]{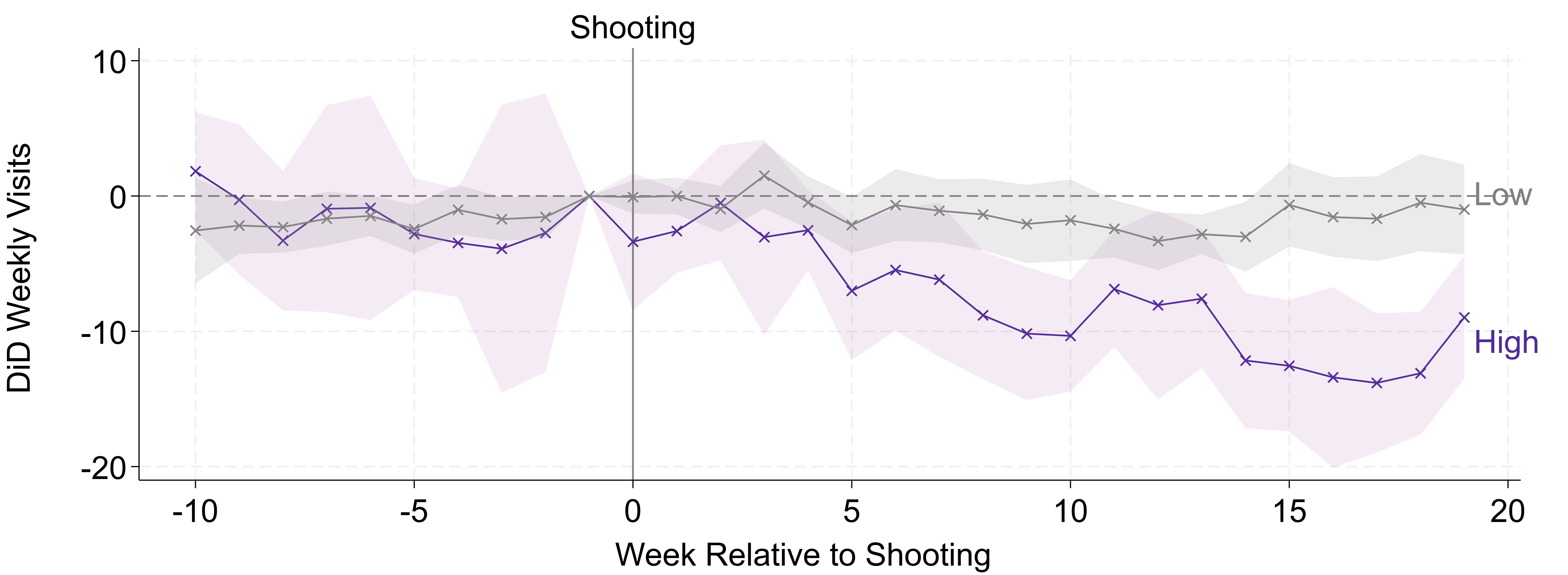}
    \end{subfigure}
    \par\bigskip
    \par\bigskip
    \begin{subfigure}{0.9\linewidth}
    \centering
        \caption{Using Competition Level}\label{substitutability_pois}
            \centering
            \includegraphics[width=0.9\textwidth]{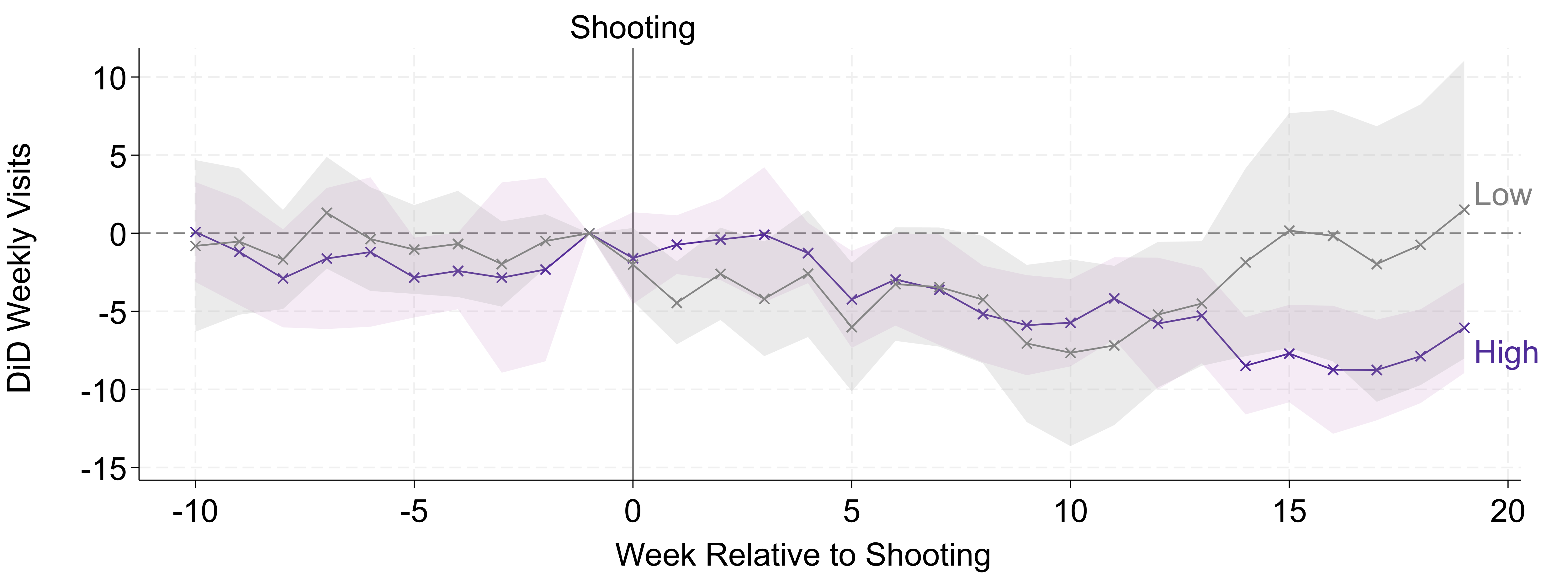}
    \end{subfigure}
    \caption*{\footnotesize \textit{*Note:} 95\% confidence intervals are shown. Standard errors are clustered at the POI level. The reference group is the POIs above median distance from the shooting site.}
\end{figure}

To generalize the idea, we evaluate how the substitutability of POIs is associated with the magnitude of the impact of mass shooting. We perform two DiD regressions, similar to that in Figure \ref{DiD}, 
for two different groups 
: \emph{high-substitutability} and \emph{low-substitutability} POIs. Substitutability is defined in two distinct ways, as shown in Table \ref{substitutability_table}. The \emph{Common Categories} measure of substitutability accounts for the nature of business based on our own experience but can be subjective. The \emph{Competition Level} measure takes into account the level of competition in a community but may be endogenous since the number of competitors or businesses is determined by local demand or differential preference among consumers.\footnote{That is, even though there are many competing locations, it may be a result of many different tastes among consumers, so substitution can still be limited.} Each measure has its pros and cons, so the results in this section should be used to gain insights, rather than quantify the difference. 

Figures \ref{substitutability_cats} and \ref{substitutability_pois} report the DiD estimates based on the \emph{Common Categories} and the \emph{Competition Level} measures, respectively. In both cases, there are distinct patterns in the evolution of foot traffic over time, with the redistribution of visits from nearby to farther-away POIs being more pronounced for highly substitutable POIs toward the end of the period. Due to the smaller sample size for each DiD regression, these differences are often not statistically significant. Nevertheless, the differential patterns are evident in these plots and the patterns are robust to an alternative specification using 4-quartile distance bands in Online Appendix D.2.

\subsection{Degree of Traumatic Stress and Magnitude of Impact}

The previous findings indicate that mass
shootings drive significant temporal and spatial displacements of economic activity. 
Given that the likelihood of another mass shooting occurring at the exact same location is nearly zero, these displacements 
should be understood in light of the psychological effects mass shootings have on communities. It has been well documented in the psychology and health literatures that traumatic events take a toll not only on survivors but also on their communities. Even those not directly affected by the event can develop post-traumatic stress symptoms, 
which is especially likely in cases of mass violence \citep{GaleaEtAl2003} and for those individuals who are physically closer to the trauma \citep{FurrEtAl2010,MattVasquez2008,MayWisco2016}. 


To examine how the degree of traumatic stress in our context influences the results, we split our sample into \emph{high-} and \emph{low-} traumatic stress groups based on three factors: preexisting crime rate, media coverage, and pre-event visit density. We then analyze how these factors affect the magnitude of the impact of mass shooting. Table \ref{mental_effect} summarizes how we split the data and predictions based on the literature. The analyses here are similar to those in Figures \ref{substitutability_cats} and \ref{substitutability_pois}, and supplemental plots are provided in Online Appendix E. 

\subsubsection{Preexisting Crime Rates}

Two theoretical frameworks from psychology inform our predictions regarding how mass shootings may have different effects depending on the preexisting levels of crime in the affected area. The first is \emph{stress inoculation}, which states that exposure to smaller instances of stress can promote resilience to future stressful episodes \citep{AshokanEtAl2016,GerberEtAl2018}. For example, \citet{AmirSol1999} find that students in an Israeli university experienced decreased levels of distress when facing a traumatic event if they had previously experienced other traumatic events. The second, \emph{desensitization}, consists of emotional numbing resulting from repeated exposure to chronic community violence \citep{KennedyCeballo2016,GaylordEtAl2011,MrugEtAl2008,NgEtAl2004}. 

It follows from these frameworks that the impact of mass shootings is expected to be greater when it occurs in a previously safer area, where the event would strike people who are less habituated to violent crime. We match each mass shooting with the homicide rate of the county where the shooting occurred in the year prior to the event. Homicide statistics are sourced from \texttt{County Health Rankings and Roadmaps}\footnote{\url{https://www.countyhealthrankings.org}}. Then, we classify the shooting areas into two groups: \textit{high-crime} areas, defined as those with above-median crime rates, and \textit{low-crime} areas, including all other locations.

Figure \ref{crime} plots the evolution of weekly visits for high- and low-crime areas. Over time, the impact of mass shootings on mobility becomes stronger when they occur in previously safer areas. That is, when communities are less habituated to violence, their reaction is stronger, resulting in more significant displacements of activities. 

\subsubsection{Extent of Media Coverage}

Evidence suggests that the psychological impact of crime is driven primarily by perceived risk, not by actual recorded crimes \citep{PearsonBreetzke2014,Braakmann2012,Hamermesh1999,Hale1996}. The media, then, would play an important role through its influence on public perception. According to \citet{MastroroccoMinale2018}, who use a staggered introduction of digital TV in Italy as a natural experiment, reduced exposure to crime news channels causes people to decrease their concerns about crime. Similarly, \citet{VelasquezEtAl2020} find that spikes in crime reporting increase people's perceptions about the prevalence of crime, even when actual crime rates decline. It follows that, all else equal, more extensive media coverage on mass shootings is likely associated with greater psychological distress at the community level, and thus stronger impacts on community mobility patterns.

\begin{figure}[h!]\caption{Change in Weekly Visits by Psychological Factors}
\centering
    \begin{subfigure}{\linewidth}
        \large
        \centering
        \renewcommand{\arraystretch}{1.25} 
            \caption{Factors Related to Psychological Effects}\label{mental_effect}
            \let\center\empty
            \let\endcenter\relax
            \centering
            \resizebox{0.9\textwidth}{!}{

\begin{tabular}{p{0.2\linewidth}|p{0.3\linewidth}p{0.1\linewidth}p{0.4\linewidth}p{0.55\linewidth}}
\hline
Factor &
  High &
  Low & Prediction & Why? \\ \hline
Preexisting Crime &
  \underline{Shootings} occurring in counties with above median homicide rates &
  Others &
  Communities respond less strongly to shootings in previously dangerous areas &
  \emph{Desensitization \& Stress Inoculation}: Kennedy and Ceballo (2016), Ashokan et al. (2016) \\
Media   \quad  \quad  Coverage &
  \underline{Shootings} receiving above-median media coverage &
  Others &
  Communities respond more strongly to shootings with high media coverage &
  \emph{Perceived Risk of Victimization}: Mastrorocco and Minale (2018), Velasquez et al (2020) \\
Baseline Visit Density &
  \underline{POIs} with above-median visits prior to the shooting &
  Others &
  Foot traffic changes more strongly for highly-visited places &
  \emph{Avoidance of Public Spaces}: Garofalo (1981), Marquet et al (2020), Navarrete-Hernandez et al (2023) \\
\hline
\end{tabular}}
    \end{subfigure}
    \par\bigskip
    \begin{subfigure}{0.73\linewidth}\caption{Preexisting Crime}\label{crime}
        \centering
        \scriptsize
 \includegraphics[width=\textwidth]{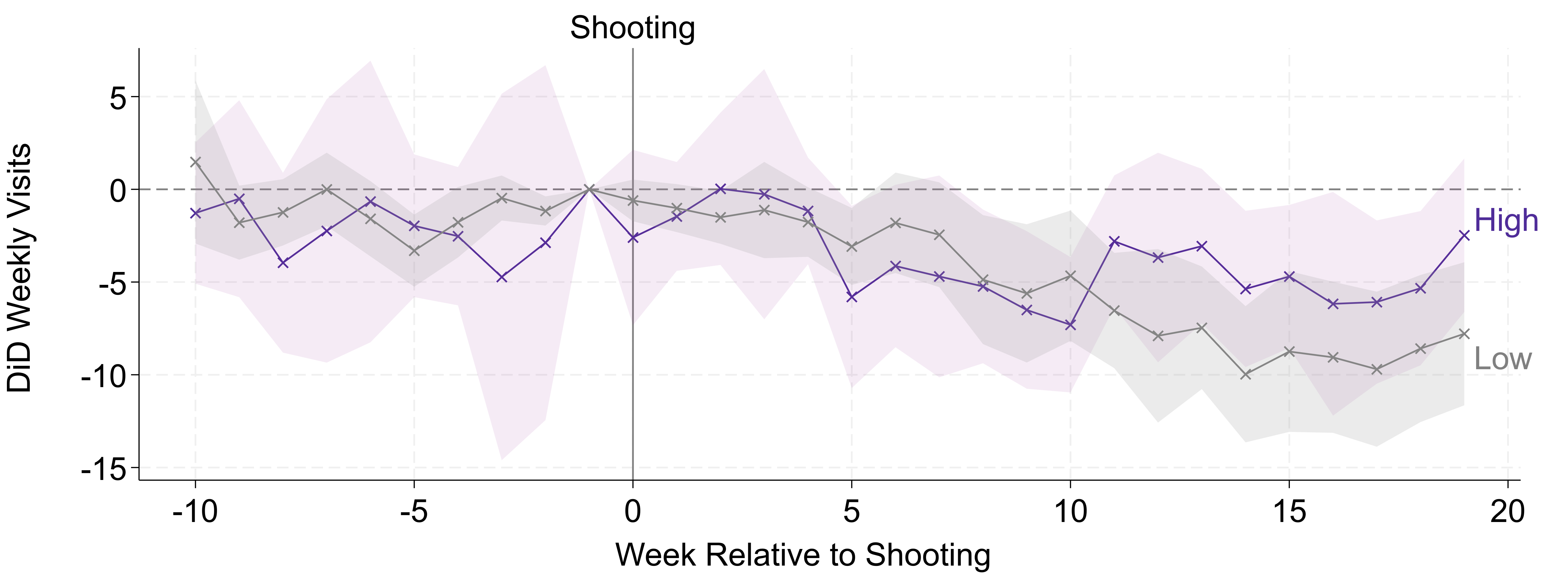}
    \end{subfigure}  
    \par\bigskip
    \begin{subfigure}{0.73\linewidth}\caption{Media Coverage}\label{media_coverage}
        \centering
        \scriptsize
        \includegraphics[width=\textwidth]{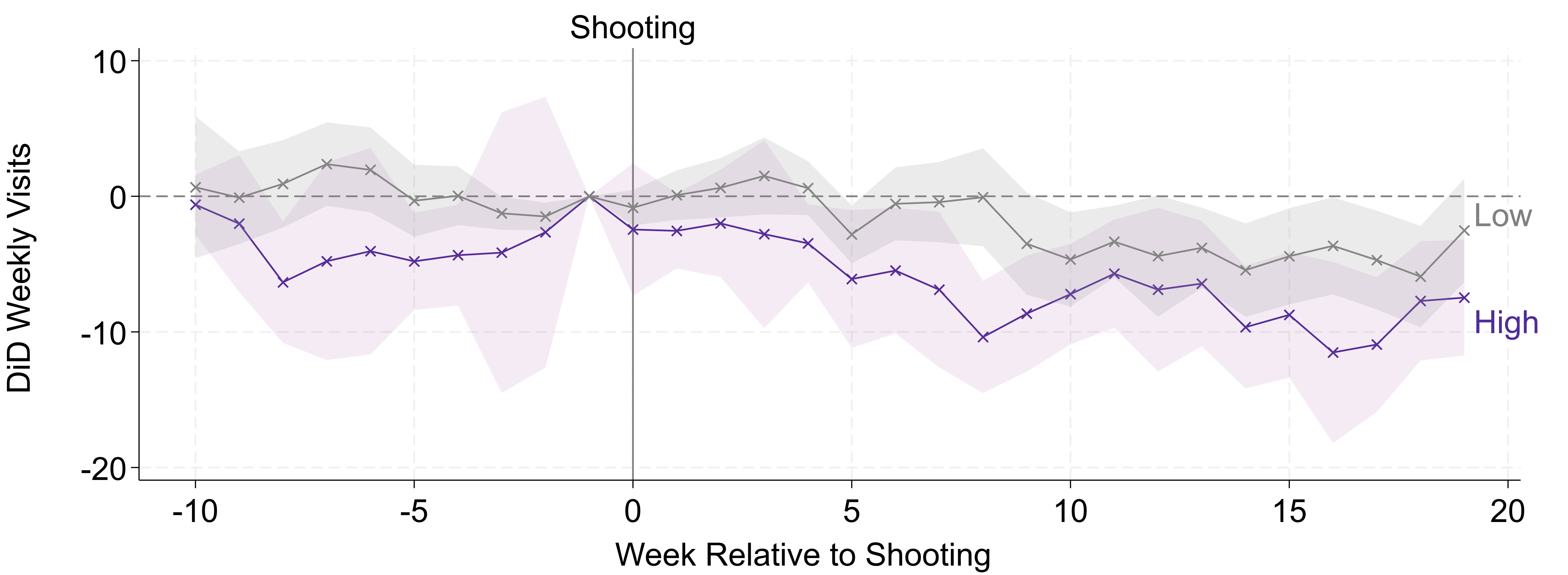}
    \end{subfigure}
    \par\bigskip
    \begin{subfigure}{0.73\linewidth}\caption{Baseline Visit Density}\label{visitor_density}
        \centering
        \scriptsize
        \includegraphics[width=\textwidth]{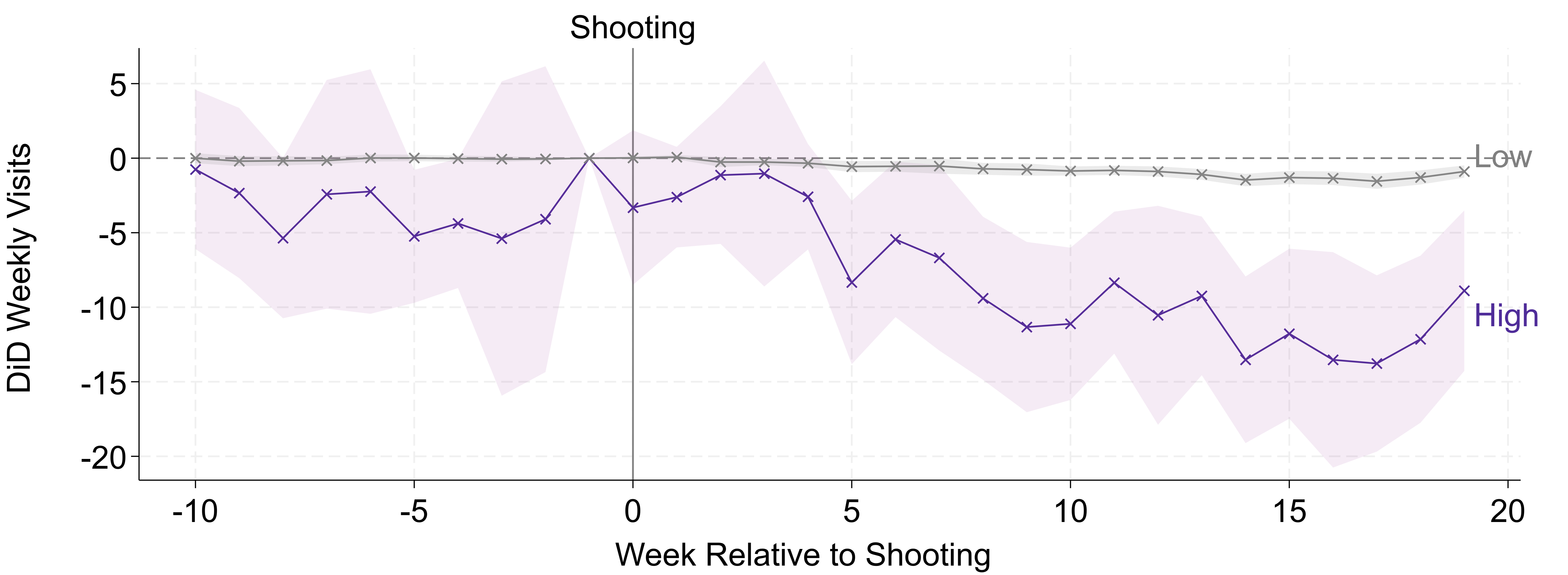}
    \end{subfigure}
    \caption*{\footnotesize \textit{*Note:} 95\% confidence intervals are shown. Standard errors are clustered at the POI level. The reference group is the POIs above median distance from the shooting site.}
\end{figure}

\clearpage
We calculate media coverage for each shooting event by the number of online news publications within one week of the event.\footnote{Number of news publications is frequently used in the literature to measure media coverage \citep[e.g.,][]{ChordiaEtAl2022}.} We identify these news publications through an internet search using broad terms associated with the shooting event and the nearby community. For example, for the Mercy Hospital shooting on November 19$^{th}$, 2018, its media coverage is the number of publications between the day of the shooting and November 26$^{th}$, 2018, identified through search terms ``Chicago shooting" and ``Illinois shooting." Our measure is a reasonable proxy for media coverage given that online news or searches are the primary  sources of news consumption for the average person today.

Shooting events are then divided in two groups: those receiving above-median media coverage (\emph{high coverage}) and those below this threshold (\emph{low coverage}). Figure \ref{media_coverage} plots the evolution of weekly visits for \emph{high-} and \emph{low-} media coverage shootings. Consistent with our prediction, more extensively publicized events tend to result in larger impacts. 

\subsubsection{Visitor Density}
Fear of crime is associated with an individual's perception of vulnerability -- that is, the self-assessed likelihood of victimization, which 
can translate into economically-relevant behavioral changes, such as the avoidance of certain physical areas or situations 
\citep{Hale1996,Garofalo1981,WhitleyPrince2005}. Furthermore, the decision to avoid a physical environment depends on the individual's assessment of potential risks, even in absence of direct connection between the location and specific instances of crime. \citep{MarquetEtAl2020, NavarreteEtAl2023}.

Public spaces, or places that are open to visitors and can, thus, attract large gatherings, are likely to be assessed as increasing the risk of victimization in the minds of the communities experiencing mass shootings. Figure \ref{visitor_density} plots the evolution of weekly visits for POIs that, prior to the shooting, received high or low levels of visits. Here, a POI is considered to have \emph{high baseline visits} if, during the week right before the shooting, it received above-median number of visits relative to all locations in the area of the shooting. All other POIs are treated as having \emph{low baseline visits}.

 The results are consistent with the prediction that communities respond more intensely for the locations where larger gatherings of people are expected, likely in an effort, conscious or not, to avoid situations that are perceived as conducive to victimization. POIs with low baseline visit density exhibit little to no change in visits after a mass shooting, with very narrow confidence intervals. The median number of nonzero weekly visits prior to the shootings among all POIs in the sample is 20, suggesting that low-baseline-visits POIs are generally very small and, as a result, do not show meaningful variation in visits.\footnote{This may indicate that the effects of mass shootings presented elsewhere in the paper can be seen as a lower bound, as they include POIs that are expected to show little change in visits due to the nature of the business. }

\section{Conclusion}\label{conclusion}


Our paper discusses an important, yet unexplored, cost of mass shootings: welfare loss due to the displacement of economic activity. Visits to nearby POIs decrease a few weeks after a mass shooting by about 26\% with respect to the baseline, while farther-away POIs experience noticeable increases. 
The impact is not only substantial in size but also long-lasting and pervasive across a variety of location types. Evidence suggests that there is relocation of economic activity, where individuals substitute away from POIs near the shooting site, particularly for places that are easily substitutable. 

A key driver of  displacement is the psychological effects of these events. The impact of mass shootings is stronger when they occur in safer areas, when they receive more attention from the media, and for POIs that previously received many visitors. Displacement of activities carries great economic implications, as it would 
involve friction costs and efficiency losses due to distortions in optimal choices of time and location. This could, in turn, negatively impact other economic outcomes, such as earnings and productivity.


\clearpage
\bibliographystyle{agsm}
\bibliography{References_massshootings}

@article{VM2025,
author= {Vachuska, Karl and Movahed, Masoud},
title={The Causal Effect of Gun Violence on Everyday Mobility Patterns Across {US} Neighborhoods},
year={2025},
journal = {Spatial Demography},
pages= {1-27},
volume = {13},
number = {1},
}

@Inbook{Yousaf2023,
author= {Yousaf, Hasin},
editor={Zimmermann, Klaus F.},
title={The Economics of Public Mass Shootings},
bookTitle= {Handbook of Labor, Human Resources and Population Economics},
year={2023},
publisher= {Springer International Publishing},
pages= {1-18}
}

@techreport{ChenPope2020,
	author = {Chen, M. Keith and Pope, Devin G},
	institution = {National Bureau of Economic Research},
	number = {27072},
	title = {`{Geographic} Mobility in {America}: Evidence from Cell Phone Data'},
	type = {Working Paper},
	year = {2020}}

@article{ChenRohla2018,
	author = {M. Keith Chen and Ryne Rohla},
	journal = {Science},
	number = {6392},
	pages = {1020-1024},
	title = {The effect of partisanship and political advertising on close family ties},
	volume = {360},
	year = {2018}}

@article{ChenEtAl2022,
	author = {Chen, M. Keith and Haggag, Kareem and Pope, Devin G. and Rohla, Ryne},
	journal = {The Review of Economics and Statistics},
	month = {11},
	number = {6},
	pages = {1341-1350},
	title = {Racial Disparities in Voting Wait Times: Evidence from Smartphone Data},
	volume = {104},
	year = {2022}}

@article{BrodeurYousaf2022,
	author = {Brodeur, Abel and Yousaf, Hasin},
	journal = {The Review of Economics and Statistics},
	pages = {1-43},
	title = {On the Economic Consequences of Mass Shootings},
	year = {2022}}

@article{DonohueEtAl2019,
	author = {Donohue, John J. and Aneja, Abhay and Weber, Kyle D.},
	journal = {Journal of Empirical Legal Studies},
	number = {2},
	pages = {198-247},
	title = {Right-to-Carry Laws and Violent Crime: A Comprehensive Assessment Using Panel Data and a State-Level Synthetic Control Analysis},
	volume = {16},
	year = {2019}}

@article{LucaEtAl2020,
	author = {Michael Luca and Deepak Malhotra and Christopher Poliquin},
	journal = {Journal of Public Economics},
        volume={181},
	number = {},
	title = {The Impact of Mass Shootings on Gun Policy},
	year = {2020}}

@article{Yousaf2021,
	author = {Yousaf, Hasin},
	journal = {Journal of the European Economic Association},
	number = {5},
	pages = {2765-2802},
	title = {Sticking to One's Guns: Mass Shootings and the Political Economy of Gun Control in the {United States}},
	volume = {19},
	year = {2021}}

@article{LoweGalea2017,
	author = {Sarah R. Lowe and Sandro Galea},
	journal = {Trauma, Violence, \& Abuse},
	number = {1},
	pages = {62-82},
	title = {The Mental Health Consequences of Mass Shootings},
	volume = {18},
	year = {2017}}

@article{AmirSol1999,
	author = {Amir, Marianne and Sol, Oren},
	journal = {Journal of Traumatic Stress},
	number = {1},
	pages = {139-154},
	title = {Psychological Impact and Prevalence of Traumatic Events in a Student Sample in {Israel}: The Effect of Multiple Traumatic Events and Physical Injury},
	volume = {12},
	year = {1999}}

@article{AshokanEtAl2016,
	author = {Ashokan, Archana and Sivasubramanian, Meenalochani and Mitra, Rupshi},
	journal = {Naural Plasticity},
	number = {},
	pages = {},
	title = {Seeding Stress Resilience through Inoculation},
	volume = {2016},
	year = {2016}}

@article{Cabral2025,
	author = {Cabral, Marika and Kim, Bokyung and Rossin-Slater, Maya and Schnell, Molly and Schwandt, Hannes},
	journal = {Review of Economic Studies},
   number = {},
	pages = {1-39},
	title = {Trauma at School: The Impacts of Shootings on Students' Human Capital and Economic Outcomes},
	volume = {00},
	year = {2025}}

@article{BelandKim2016,
	author = {Beland, Louis-Philippe and Kim, Dongwoo},
	journal = {Educational Evaluation and Policy Analysis},
	number = {1},
	pages = {113-126},
	title = {The Effect of High School Shootings on Schools and Student Performance},
	volume = {38},
	year = {2016}}

@article{VelasquezEtAl2020,
	author = {Velasquez, Daniel and Medina, Santiago and Yamada, Gustavo and Lavado, Pablo and Nunez-del-Prado, Miguel and Alatrista-Salas, Hugo and Morzan, Juandiego},
	journal = {World Development},
	number = {136},
	pages = {},
	title = {I Read the News Today, Oh Boy: The Effect of Crime News Coverage on
Crime Perception},
	volume = {},
	year = {2020}}

@article{SharkeyShen2021,
	author = {Sharkey, Patrick and Shen, Yinzhi},
	journal = {PNAS},
	number = {23},
	pages = {},
	title = {The Effect of Mass Shootings on Daily Emotions is Limited by Time, Geographic Proximity, and Political Affiliation},
	volume = {118},
	year = {2021}}

@article{WozniakEtAl2020,
	author = {Wozniak, Jana DeSimone and Caudle, Hailey E. and Harding, Kaitlin and Vieselmeyer, Julie and Mezulis, Amy H.},
	journal = {Psychological Trauma: Theory, Research, Practice, and Policy},
	number = {3},
	pages = {227-234},
	title = {The Effect of Trauma Proximity and Ruminative Response Styles on Posttraumatic Stress and Posttraumatic Growth Following a University Shooting},
	volume = {12},
	year = {2020}}

@article{NavarreteEtAl2023,
	author = {Navarrete-Hernandez, Pablo and Luneke, Alejandra and Truffello, Ricardo and Fuentes, Luis},
	journal = {Landscape and Urban Planning},
	number = {237},
	pages = {},
	title = {Planning for Fear of Crime Reduction: Assessing the Impact of Public Space Regeneration on Safety Perceptions in Deprived Neighborhoods},
	volume = {},
	year = {2023}}

@article{LevineMcKnight2024,
	author = {Levine, Phillip B. and McKnight, Robin},
	journal = {Journal of Policy Analysis and Management},
	number = {43},
	pages = {1034-1056},
	title = {The Consequences of High-Fatality School Shootings for Surviving Students},
	volume = {},
	year = {2024}}

@article{MastroroccoMinale2018,
	author = {Mastrorocco, Nicola and Minale, Luigi},
	journal = {Journal of Public Economics},
	number = {165},
	pages = {230-255},
	title = {News Media and Crime Perceptions: Evidence from a Natural Experiment},
	volume = {},
	year = {2018}}

@article{Garofalo1981,
	author = {Garofalo, James},
	journal = {Journal of Criminal Law and Criminology},
	number = {2},
	pages = {839-857},
	title = {The Fear of Crime: Causes and Consequences},
	volume = {72},
	year = {1981}}

@article{GerberEtAl2018,
	author = {Gerber, Monica M. and Frankfurt, Sheila B. and Contractor, Ateka A. and Oudshoorn, Kelsey and Dranger, Paula and Brown, Lily A.},
	journal = {Journal of Psychopathology and Behavioral Assessment},
	number = {40},
	pages = {645-654},
	title = {Influence of Multiple Traumatic Event Types on Mental Health Outcomes: Does Count Matter?},
	volume = {},
	year = {2018}}

@article{MarquetEtAl2020,
	author = {Marquet, Oriol and Ogletree, S. Scott and Hipp, J. Aaron and Suau, Luis J. and Horvath, Candice B. and Sinykin, Alexander and Floyd, Myron F.},
	journal = {Preventing Chronic Disease: Public Health Research, Practice, and Policy},
	number = {},
	pages = {},
	title = {Effects of Crime Type and Location on Park Use Behavior},
	volume = {17},
	year = {2020}}

@article{Hale1996,
	author = {Hale, C.},
	journal = {International Review of Victimology},
	number = {},
	pages = {79-150},
	title = {Fear of Crime: A Review of the Literature},
	volume = {4},
	year = {1996}}

@article{GaylordEtAl2011,
	author = {Gaylord-Harden, N. K. and Cunningham, J. A. and Zelencik, B.},
	journal = {Journal of Abnormal Child Psychology},
	number = {},
	pages = {711-719},
	title = {Effects of Exposure to Community Violence on Internalizing Symptoms: Does Desensitization to Violence Occur in {African American} Youth?},
	volume = {39},
	year = {2011}}

@article{MrugEtAl2008,
	author = {Mrug, S. and Loosier, P. S. and Windle, M.},
	journal = {American Journal of Orthopsychiatry},
	number = {},
	pages = {70-84},
	title = {Violence Exposure Across Multiple Contexts: Individual and Joint Effects on Adjustment},
	volume = {78},
	year = {2008}}

@article{NgEtAl2004,
	author = {Ng-Mak, D. S. and Salzinger, S. and Feldman, R. S. and Stueve, C. A.},
	journal = {American Journal of Orthopsychiatry},
	number = {},
	pages = {196-208},
	title = {Pathologic adaptation to community violence among inner-city youth},
	volume = {74},
	year = {2004}}

@article{KennedyCeballo2016,
	author = {Kennedy, Traci M. and Ceballo, Rosario},
	journal = {Developmental Psychology},
	number = {5},
	pages = {778-789},
	title = {Emotionally Numb: Desensitization to Community Violence Exposure Among Urban Youth},
	volume = {52},
	year = {2016}}

@article{Braakmann2012,
	author = {Braakman, Nils},
	journal = {Journal of Economic Behavior \& Organization},
	number = {},
	pages = {335-344},
	title = {How do individuals deal with victimization and victimization risk? {Longitudinal} evidence from {Mexico}},
	volume = {84},
	year = {2012}}

@article{Hamermesh1999,
	author = {Hamermesh, D. S.},
	journal = {Journal of Urban Economics},
	number = {2},
	pages = {311–330},
	title = {Crime and the Timing of Work},
	volume = {45},
	year = {1999}}

@article{DustmannFasani2014,
	author = {Dustmann, Christian and Fasani, Francesco},
	journal = {The Economic Journal},
	number = {},
	pages = {978-1017},
	title = {The Effect of Local Area Crime on Mental Health},
	volume = {126},
	year = {2014}}

@article{WhitleyPrince2005,
	author = {Whitley, Rob and Prince, Martin},
	journal = {Social Science and Medicine},
	number = {},
	pages = {1678-1688},
	title = {Fear of crime, mobility and mental health in inner-city {London, UK}},
	volume = {61},
	year = {2005}}

@article{PearsonBreetzke2014,
	author = {Pearson, Amber L. and Breetzke, Gregory D.},
	journal = {Social Indicators Research},
	number = {},
	pages = {281-294},
	title = {The Association Between the Fear of Crime, and Mental
and Physical Wellbeing in {New Zealand}},
	volume = {119},
	year = {2014}}

@article{GuiteEtAl2006,
	author = {Guite, H.F. and Clark, C. and Ackrill, G.},
	journal = {Public Health},
	number = {},
	pages = {1117-1126},
	title = {The impact of the physical and urban environment on mental well-being},
	volume = {120},
	year = {2006}}

@article{Klemperer1987,
	author = {Klemperer, Paul},
	journal = {The Quarterly Journal of Economics},
	number = {2},
	pages = {375-394},
	title = {Markets with Consumer Switching Costs},
	volume = {102},
	year = {1987}}

@article{Wilson2012,
	author = {Wilson, Chris M.},
	journal = {European Economic Review},
	number = {56},
	pages = {1070-1086},
	title = {Market Frictions: A Unified Model of Search Costs and Switching Costs},
	volume = {},
	year = {2012}}

@article{Wezsacker1984,
	author = {Weizsäcker, C. Christian von},
	journal = {Econometrica},
	number = {5},
	pages = {1085-1116},
	title = {The Costs of Substitution},
	volume = {52},
	year = {1984}}

@article{GaleaEtAl2003,
	author = {Galea, Sandro and Vlahov, David and Resnick, Heidi and Ahern, Jennifer and Susser, Ezra and Gold, Joel and Bucuvalas, Michael and Kilpatrick, Dean},
	journal = {American Journal of Epidemiology},
	number = {6},
	pages = {514-524},
	title = {Trends of Probable Post-Traumatic Stress Disorder in {New York City} after the {September} 11 Terrorist Attacks},
	volume = {158},
	year = {2003}}

@article{MattVasquez2008,
	author = {Matt, Geor E. and Vasquez, Carmelo},
	journal = {The Spanish Journal of Psychology},
	number = {2},
	pages = {503-515},
	title = {Anxiety, Depressed Mood, Self-Esteem, and Traumatic Stress Symptoms among Distant Witnesses of the 9/11 Terrorist Attacks: Transitory Responses and Psychological Resilience},
	volume = {11},
	year = {2008}}

@article{MayWisco2016,
	author = {May, Casey L. and Wisco, Blair E.},
	journal = {Psychological Trauma: Theory, Research, Practice, and Policy},
	number = {2},
	pages = {233-240},
	title = {Defining Trauma: How Level of Exposure and Proximity Affect Risk for Posttraumatic Stress Disorder},
	volume = {8},
	year = {2016}}

@article{FurrEtAl2010,
	author = {Furr, Jami M. and Comer, Jonathan S. and Edmunds, Julie M. and Kendall, Philip C.},
	journal = {Journal of Consulting and Clinical Psychology},
	number = {6},
	pages = {765-780},
	title = {Disasters and Youth: A Meta-Analytic Examination of Posttraumatic Stress},
	volume = {78},
	year = {2010}}

@article{BharadwajEtAl2021,
        author = {Bharadwaj, Prashant and Bhuller, Manudeep and Løken, Katrine V. and Wentzel, Mirjam},
        year = {2021},
        journal = {Journal of Public Economics},
        title = {Surviving a mass shooting},
        volume = {200},
        number = {},
        pages = {}
        }

@article{ChordiaEtAl2022,
        author = {Chordia, Tarum and Jeung, Jinoug and Pati, Abinash},
        year = {2022},
        journal = {Available at SSRN: https://ssrn.com/abstract=4269600},
        title={The Price of Tragedy: Mass Shootings, Salience Bias, and Municipal Bond Yields},
        volume = {},
        number = {},
        pages = {}
        }

@article{GarciaCowan2024,
        author = {Garcia, K.S.D. and Cowan, B.W.},
        year = {2024},
        journal = {Journal of Labor Research},
        title={Childcare Responsibilities and Parental Labor Market Outcomes During the {COVID-19} Pandemic},
        volume = {45},
        number = {},
        pages = {153-200}
        }

\end{document}